\documentclass{trbunofficial}
\usepackage{graphicx}
\usepackage{amsmath,amsfonts}
\usepackage{color}

\usepackage{hyperref}

\newlength\savewidth\newcommand\shline{\noalign{\global\savewidth\arrayrulewidth
		\global\arrayrulewidth 1pt}\hline\noalign{\global\arrayrulewidth\savewidth}}

\AuthorHeaders{Meng, Sayer, Zhang, Shen, Li, and Liu}
\title{ROCO: A Roundabout Traffic Conflict Dataset}

\author{%
  \textbf{Depu Meng\footnote{This work was done when
  Depu Meng was a visiting scholar at the Department of Civil and Environmental Engineering, University of Michigan, Ann Arbor.  }}\\
  Department of Automation \\
  University of Science and Technology
  of China, Hefei, China \\
  Email: \texttt{mdp@mail.ustc.edu.cn} \\
  \hfill\break
  \textbf{Owen Sayer}\\
  Department of Mechanical Engineering \\
  University of Michigan, Ann Arbor, Michigan \\
  Email: \texttt{osayer@umich.edu} \\
  \hfill\break%
  \textbf{Rusheng Zhang}\\
   Department of Civil and Environmental Engineering \\
   University of Michigan, Ann Arbor, Michigan \\
   Email: \texttt{rushengz@umich.edu} \\
   \hfill\break%
   \textbf{Shengyin Shen}\\
   University of Michigan Transportation Research Institute \\
   University of Michigan, Ann Arbor, Michigan \\
   Email: \texttt{shengyin@umich.edu} \\
   \hfill\break%
  \textbf{Houqiang Li, Ph.D.}\\
  Department of Electronic Engineering and Information Science \\
  University of Science and Technology of China,
  Hefei, China \\
  Email: \texttt{lihq@ustc.edu.cn} \\
  \hfill\break%
  \textbf{Henry X. Liu, Ph.D., Corresponding Author}\\
  Department of Civil and Environmental Engineering \\
  University of Michigan, Ann Arbor, Michigan \\
  Email: \texttt{henryliu@umich.edu}
}

\begin{document}
\maketitle

\section{Abstract}
Traffic conflicts have been studied by the transportation research community
as a surrogate safety measure for decades.
However, due to the
rarity of traffic conflicts, 
collecting large-scale real-world traffic conflict data
becomes extremely challenging.
In this paper, we introduce and analyze ROCO - a real-world
roundabout traffic conflict dataset.
The data is collected at a two-lane roundabout at the intersection
of State St. and W. Ellsworth Rd. in Ann Arbor, Michigan.
We use raw video dataflow captured
from \textit{four} fisheye cameras installed at the roundabout
as our input data source.
We adopt a learning-based conflict identification
algorithm from video to find potential traffic conflicts,
and then manually label them for dataset collection and annotation.
In total $557$ traffic conflicts and $17$ traffic crashes
are collected from August $2021$ to October $2021$.
We provide 
trajectory data of the traffic conflict scenes extracted using our
roadside perception system. Taxonomy based on traffic conflict
severity, reason for the traffic conflict, and its effect on the
traffic flow is provided. With the traffic conflict data
collected, we discover that
failure to yield to circulating
vehicles when entering the roundabout is the largest contributing
reason for traffic conflicts.
ROCO dataset will be made public at \url{https://github.com/michigan-traffic-lab/ROCO}.

\hfill\break%
\noindent\textit{Keywords}: traffic conflict, safety measure,
roadside perception, safety analysis
\newpage

\section{Introduction}

Improving traffic safety has been an essential and longstanding
goal in transportation research for decades.
Traffic accidents like crashes, are the dominant
cause of fatalities, injuries, and property damage
in the transportation sector. 
For better understanding of car crashes, large-scale crash data is
required.
However, due to the rarity of crash occurrences ~\cite{LiuF2022,feng2021intelligent} (around 
$1.9\times 10^{-6}$ per mile),
and the very limited data collection methods
(in most cases, crash data are collected from police reports),
it is very hard to acquire large amounts of crash data
in a relatively short period. Additionally, the information contained in police reports often lacks necessary information, \emph{e.g.,} exact vehicle speed and
maneuver before the crash happens.

\begin{figure}[t!]
    \centering
    (a)~\includegraphics[width=.95\textwidth]{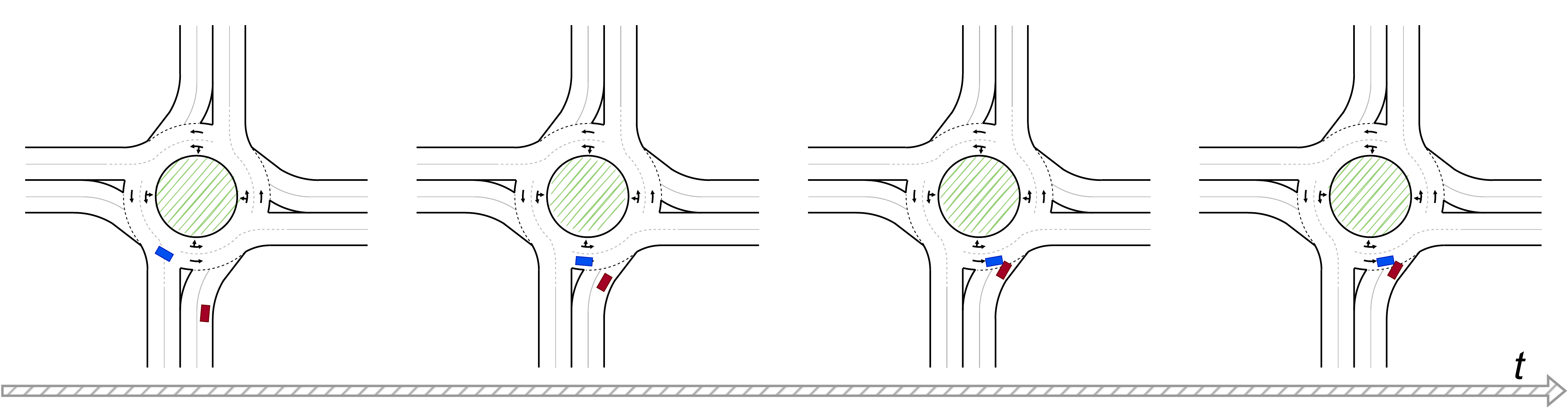}\\
    (b)~\includegraphics[width=.95\textwidth]{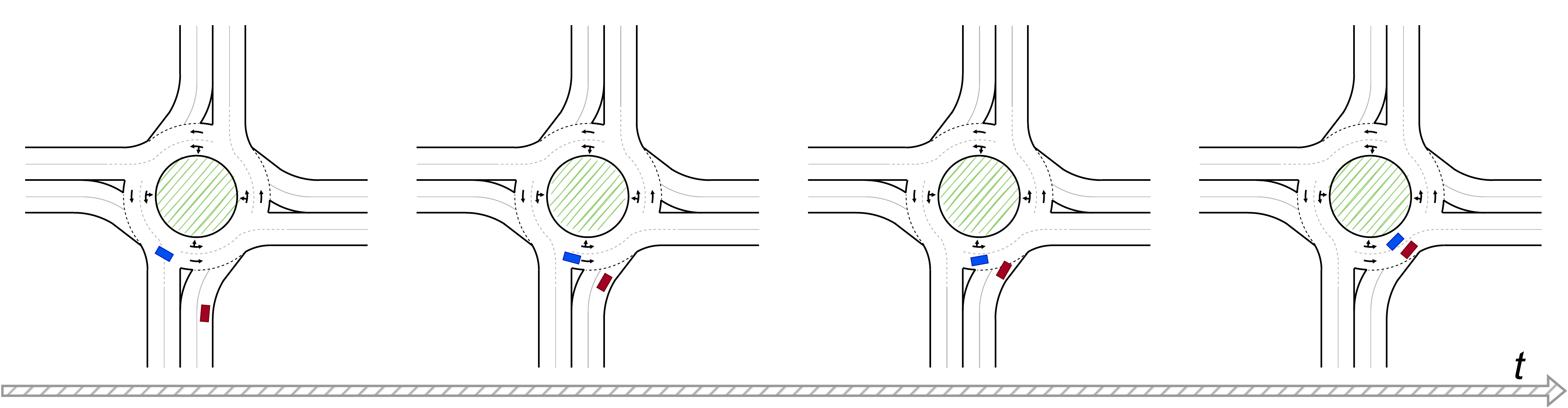}\\
    (c)~\includegraphics[width=.95\textwidth]{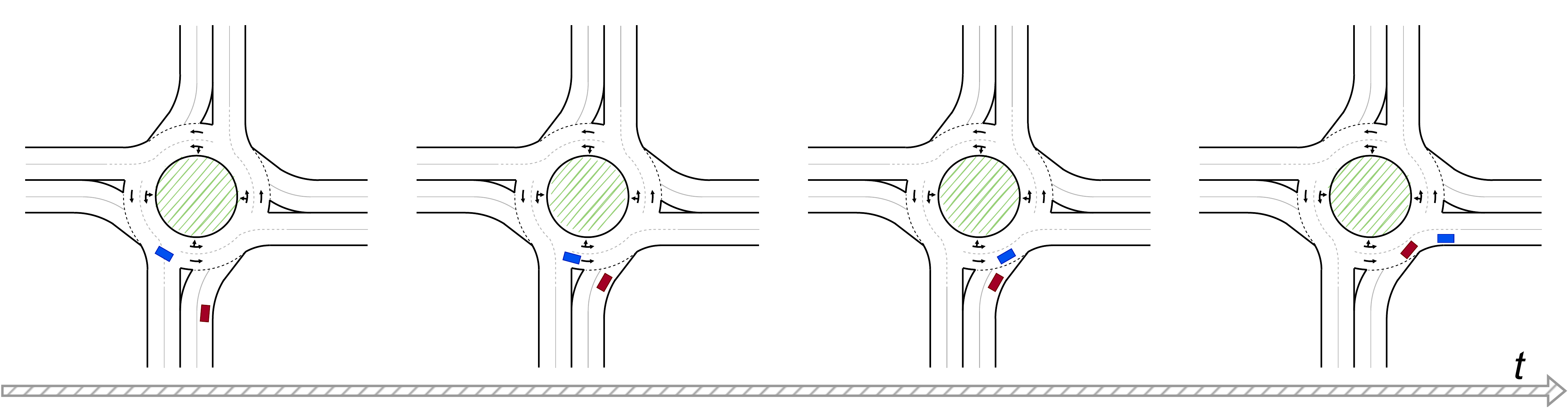}\\
    \caption{Illustrations of \textbf{(a)} crashes,
    \textbf{(b)} traffic conflicts and \textbf{(c)}
    normal driving behaviors.
    In (a), the red vehicle enters the roundabout
    without yielding to the blue vehicle and causes a crash.
    In (b), the red vehicle fails
    to yield and forces the blue vehicle to take
    an evasive action (sharply change to the inner lane)
    and causes a traffic conflict.
    In (c), the red vehicle yields the blue vehicle and
    no crash or conflict happens.
    A traffic conflict is considered a broad definition that
    includes temporal-spatial proximity or evasive actions,
    near-misses, near-crashes, and other safety-critical events 
    ~\cite{ZHENG2021100142}.
    Traffic conflict can serve as a surrogate safety measure
    to traffic accidents to help transportation researchers
    and engineers better improving traffic safety.}
    \label{fig:teaser}
\end{figure}

Due to the limitations as mentioned
in crash data acquisition,
surrogate safety measures
like traffic conflict have been widely studied and discussed.
The term \emph{traffic conflict} has no universal definition.
In this paper, we use the definition given by
NCHRP 219~\cite{glauz1980application}: "A traffic conflict is a traffic event
involving two or more road users, 
in which one user performs some atypical
or unusual action, such as a change in direction
or speed, that places another user in jeopardy of
collision unless an evasive maneuver is undertaken." 
Figure~\ref{fig:teaser} shows a comparison
between crashes and traffic conflicts.
The traffic conflicts hold as the "missing link"
between normal traffic behavior and accidents~\cite{cooper1977}.
Many studies on the relationship
between traffic conflicts and crashes have been conducted,
including pyramid safety continuum~\cite{hyden1987development}, 
aetiology consistency~\cite{tarko2018estimating},
and causal model~\cite{davis2011outline}.
Since traffic conflicts occur more frequently,
and typically no damage takes place,
it is easier to acquire large amounts of traffic
conflict data in smaller amounts of time.
A variety of traffic conflict applications have been studied,
including critical behavior analysis
~\cite{madsen2017comparison,beitel2018assessing,MA2018303}, before-after studies
~\cite{brow1994traffic,Tarrall1998,Ismail2010},
identifying high-risk locations~\cite{TIWARI1998207},
and real-time safety
prediction
~\cite{HOSSAIN201966,machiani2016safety,ESSA2018289}.

Though traffic conflict data is easier to acquire
than crash data, several data-related issues still
exist in traffic conflict research.
One of the most critical issues is the lack of publicly
available large-scale traffic conflict data.
As it stands, transportation researchers and engineers
need to collect traffic conflict data each themselves.
Furthermore, since every research group uses their own
data, as mentioned in~\cite{ZHENG2021100142},
data inconsistency becomes an issue.
Different conflict identification criteria lead to
different analysis results, and it is difficult to discuss and generalize
findings from different research because of this.
Most traffic conflict data is collected through
a roadside camera, or connected vehicles.
Some factors that potentially contribute to traffic
conflicts, such as the state of other surrounding vehicles,
weather conditions, and lighting conditions
may be unobserved in traffic conflict data.
These unobserved factors might cause inaccurate
modeling and analysis of traffic safety.

\begin{figure}
    \centering
    (a)~\includegraphics[height=.27\textwidth]{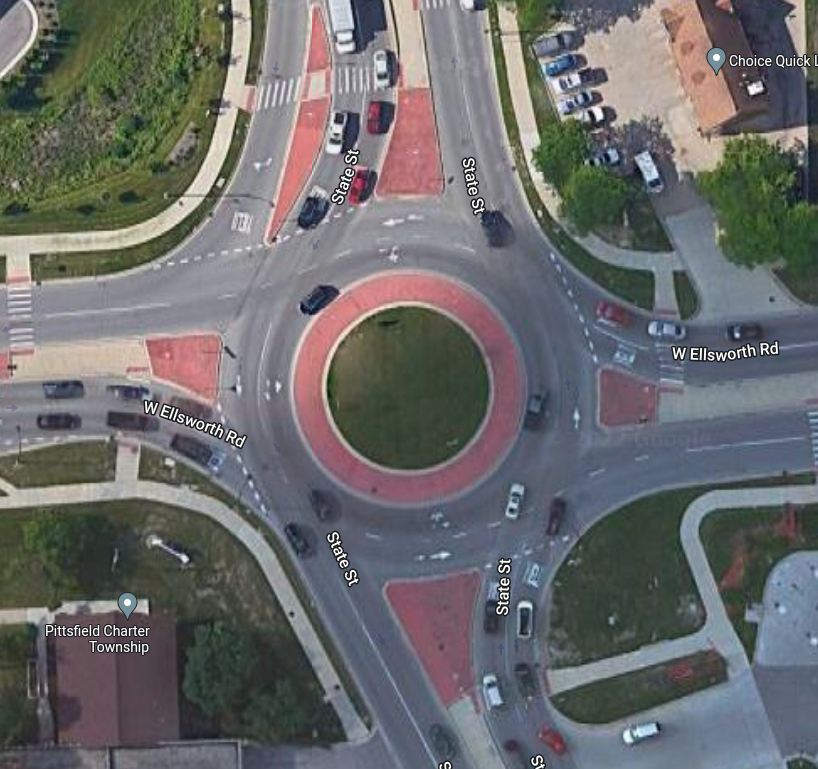}~~~
    (b)~\includegraphics[height=.27\textwidth]{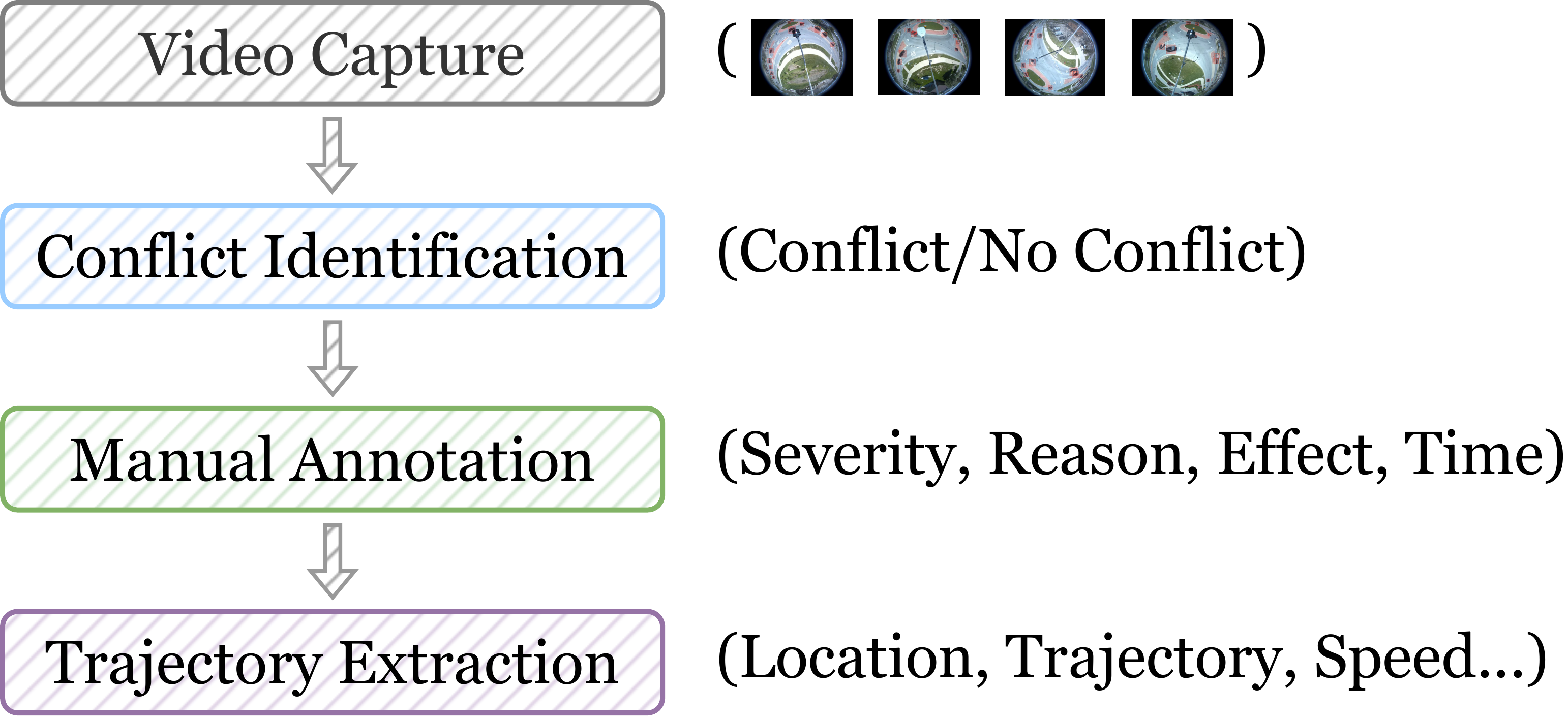}
    \caption{Illustrations of (a) the roundabout
    where data is collected. The roundabout is a
    two-lane roundabout located at the intersection of State St.
    and W. Ellsworth Rd. in Ann Arbor, Michigan, U.S.
    (b) traffic conflict data
    collection pipeline. We first use a learning-based
    conflict identification algorithm to find the potential
    conflicts, then we manually check and label the potential
    conflicts and extract the trajectory data.}
    \label{fig:data-collection-pipeline}
\end{figure}

To overcome the aforementioned traffic conflict data issues,
we propose ROCO, a new \underline{ro}undabout traffic \underline{co}nflict
dataset.
In order to collect rarely occurring traffic conflict
data, we utilize pre-installed roadside
cameras and an edge-cloud infrastructure 
\footnote{Mcity sponsored
research project N029385: Safe and Efficient
Roundabouts by Artificial Intelligence and V2X Technology.}.
Four roadside fisheye
cameras were installed on the four corners
of the 
roundabout.
Meanwhile, an edge-cloud infrastructure is deployed to stream the raw sensor data gathered from the roadside to the cloud-based
Mcity Mobility Data Center in a 24/7 manner with low latency.~\cite{Zhang2021,Zou2021},
as shown in Figure~\ref{fig:data-collection-pipeline}(a).
The aforementioned roundabout where the ROCO data is collected is a two-lane roundabout
at the intersection of State St. and W. Ellsworth Rd. ($42.229379$, $-83.739013$) in Ann Arbor, Michigan, U.S.
Even with the $24/7$ video data collected,
it is too time intensive to identify the traffic conflicts
from normal traffic scenarios manually.
Therefore, we developed a data collection
and annotation pipeline that
first uses a deep learning-based video processing algorithm
for potential traffic conflict identification,
and then we manually check and label all
the video clips that were flagged as potentially containing traffic conflicts.
The traffic conflict identification algorithm
is a two-stage architecture. A modified SlowFast
~\cite{feichtenhofer2019slowfast}
network is adopted for each stage.
The first-stage model identifies a significant portion of
the video clips.
The second-stage model performs a re-identification
among the videos filtered from the first stage. After the second stage, about $23\%$ of filtered
videos contain one or more traffic conflicts.

With the video clips filtered by the two-stage algorithm, we begin the verification
and annotation work.
We do not provide a quantitative definition
for traffic conflicts, because the traffic conflict
is a multi-dimensional problem
~\cite{VOGEL2003427,Guido2011481,laureshyn2017search}.
It is very hard to provide a generalized
explicit quantitative condition for traffic conflicts.
Instead, we use exemplar-based annotation.
We provide a few examples of each typical type
of traffic conflict in the roundabout,
and annotators qualitatively estimate
the similarity between a potential conflict and existing exemplars.
We design a taxonomy of traffic conflicts in terms of conflict severity, the reason for the traffic conflict, and the effect of the traffic conflict on the
traffic flow in the roundabout.
There are in total $557$ traffic conflict
samples and $17$ traffic crash samples
in the ROCO dataset. Further, we use a roadside perception system to extract the trajectory data from each video clip. With the full sensor coverage of the roundabout, all vehicles inside or near the roundabout are
recorded. Lighting and
weather conditions are also recorded for convenience in future research.

Some preliminary statistical results
of the ROCO dataset are provided in this paper.
One interesting finding is that
the biggest reason for traffic conflicts
is failure to yield to circulating vehicles
when entering the roundabout.
More than $71\%$ of the traffic conflicts
we collected are caused by failure to yield.
Further, although large vehicles, like trucks,
are not very common in the roundabout,
they cause a lot of the failure to yield cases
(about $1/3$ of all failure-to-yield cases
are caused by large vehicles).

We believe that with ROCO dataset,
transportation researchers will be
able to easily access consistently labeled
real-world traffic conflict data.
Modeling traffic conflicts, as well as studying
the relationship between crashes and conflicts,
will provide guidance on how to improve
roundabout traffic safety.
Currently, we only collect data from
one roundabout. In the future, however,
we will continue to collect more traffic
conflict data from multiple roundabouts
and signalized intersections
to improve the diversity and generalization
of the dataset.

In summary, the main contributions of
this paper are as follows:
\begin{enumerate}
    \item We present ROCO, a roundabout
    conflict dataset with rich contextual information,
    including surrounding vehicle states and
    weather conditions.
    \item We design a data collection pipeline
    for traffic conflicts, including
    a two-stage conflict identification
    framework and human verification and annotation.
    \item We perform statistical analysis
    on the ROCO dataset and discover patterns
    in traffic conflicts so that transportation
    agencies can take
    to improve roundabout safety and efficiency.
\end{enumerate}

\section{Related Work}
Research on traffic safety analysis, especially the automatic detection of traffic conflicts, is an ever-progressing field. As this research progresses, so do the techniques and approaches which build upon the work of those before them. Just as traffic conflicts were first theorized, they can then be validated as a crash-predictive metric and used to analyze the roads around us.

\subsection{Traffic conflict}
The very first roots of traffic conflicts were envisioned in the 1930's when Herbert W. Heinrich proposed that all safety-related accidents share common causes or origins with near-misses~\cite{article3}. This idea can be specifically applied to traffic conflicts and crashes by assuming that a crash is a potential outcome from the initial state of a conflict (one in which two or more vehicles will collide should no evasive action be taken)~\cite{ZHENG2021100142,ORSINI2021106382}. 
This definition has since been expanded on.
Their use in traffic analysis has also increased as the problems with crash-based analysis are further realized. 

Traffic conflicts are used as a surrogate to crash data for a number of reasons. Traffic conflicts happen much more often than crashes do~\cite{ORSINI2021106382,Glauz1985EXPECTEDTC}, although they are harder to identify; a crash is clearly defined by contact between two or more vehicles, but the identification of a traffic conflict is much more subjective~\cite{ZHENG2021100142}. However, they still allow us to find and utilize a much larger and more diverse set of data in traffic analyses. Another reason that traffic conflicts are used rather than crash data is that crash data is "inherently reactive"~\cite{ZHENG2021100142,article}, to obtain crash data a crash must occur. This can significantly delay improvements and changes to traffic control that prevent injuries and deaths. Additionally, crash data is often unreliable or not widely available~\cite{ZHENG2021100142,article}. The data is often collected from police reports, which do not always include important factors like the exact location of the crash or accurate vehicle speeds. Police reports also do not account for all crashes, as many go unreported~\cite{article2}.

Traffic conflicts are believed to serve as an accurate or reasonable crash precursor~\cite{ORSINI2021106382}. If traffic conflicts occur when one or more road user performs an evasive action to avoid a collision or crash, then it logically follows that the failure of this evasive action would result in a crash. This view is laid out by a number of different papers and models. This includes the pyramid safety continuum model which ranks all traffic events on how dangerous they are; they fall into three categories, undisturbed passages, traffic conflicts, and crashes~\cite{ZHENG2021100142,ZHENG2014155}. As well as in, among many others,~\cite{ELBASYOUNY2013160} and~\cite{7795980}, which both found a strong correlation between traffic conflicts and crashes in their respective studies. 

Traffic conflicts are typically determined either by eye and a set of criteria that a study or group establishes, or by using a crash-predictive metric and an algorithm. Such metrics include TTC (time to collision) and PET (post-encroachment time). TTC is the amount of time before two or more vehicles collide should all conditions remain the same. TTC has been used widely, such as in ~\cite{ORSINI2021106382} and \cite{ELBASYOUNY2013160}. PET is the time after one vehicle has left a potential collision point and the time that the other vehicle arrives there ~\cite{ZHENG2021100142}. By setting a threshold value for TTC or PET, conflicts can be automatically detected.

Despite the clear benefits of traffic conflict analyses, traffic conflicts suffer from data and collection problems. First and foremost, there are no universal standards for traffic conflict identification~\cite{ZHENG2021100142}. This results in inconsistencies across studies. For conflicts detected by the algorithm, there must be a set of rules or standards that guide the computers decision, but these are determined nearly entirely by the researchers themselves~\cite{ZHENG2021100142}. Because of this, comparing the results of different studies and validating results becomes extremely challenging. Additionally, traffic conflict data, like crash data, is often collected in a short period of time. Data collection of just a few days or even weeks lacks the ability to account for seasonal, event-based, or weather-based changes in driver behavior and traffic volumes~\cite{NOFAL199751}. Which leads to biased assumptions of traffic conflict volumes across various times~\cite{ZHENG2021100142}. Understanding these flaws is key to seeing why more standardized data is necessary. 

By setting specific standards or assembling a large pool of consistent data, as is done in this study, we give researchers access to data that can easily be compared and referenced against each other. This allows models to be validated and cross-referenced and overall improved. It additionally allows us to take a closer look at other variables that are often ignored or considered less important. These include lighting, weather, time of day, lane in the intersection or roundabout, etc.

\subsection{Roadside perception}
Roadside perception systems are an essential
component in intelligent transportation systems,
connected vehicles, and autonomous vehicles
~\cite{chen2019cooper,rauch2012car2x,wang2020v2vnet}.
Compared to vehicle-side perception systems,
roadside perception systems
are more robust to occlusions and better at
long-term event tracking and prediction~\cite{Ye2022rope}.
~\cite{Ye2022rope} proposes a large-scale
roadside $3$d object detection dataset.
A large-scale vehicle-infrastructure cooperative 3D object detection
dataset is proposed in~\cite{yu2022dairv2x}.
A roadside perception unit (RSPU) is developed
by~\cite{tsukada2020networked}.
~\cite{Zhang2021,Zou2021} design and deploy a roadside
perception system at a roundabout.
Roadside cameras are also used in traffic conflict
research~\cite{AUTEY2012120}.
In most cases, object locations and trajectories
are extracted from the roadside perception system
and then used for traffic conflicts.
However, we directly use raw video
as our input instead of trajectories.
In our observation, the noise
in the extracted trajectories has
a negative impact on the traffic conflict
identification.
When we identified the traffic conflicts,
we used the perception system for
traffic conflict trajectory extraction.

\section{Traffic Conflict Identification}
In this section,
we introduce how we identify and collect conflict data.
As traffic conflicts are rare events,
identifying traffic conflicts purely
by hand is impractical.
On the other hand, due to the fact
that it is hard to
give a quantitative description of traffic conflicts
(compared with collisions, which we can define with physical contact between road users),
purely relying on the algorithm
to identify traffic conflicts is also not reliable. 
Therefore, we implement a traffic conflict
identification algorithm to find
potential traffic conflicts
and then manually check them.

\subsection{Input data}
We first covert the video streams
to short video clips, which are each $30$ seconds
in length with a sample rate $2.5$Hz.
The resolution of the videos is set to $960\times 720$.
Due to the distortion of the fisheye cameras,
vehicles at the opposite corner of the
roundabout become tiny and unclear in the video.
Therefore, we use the videos from
four fisheye cameras independently
instead of putting four videos from four cameras
together for conflict identification.

\begin{figure}[t]
    \centering
    \includegraphics[width=.95\textwidth]{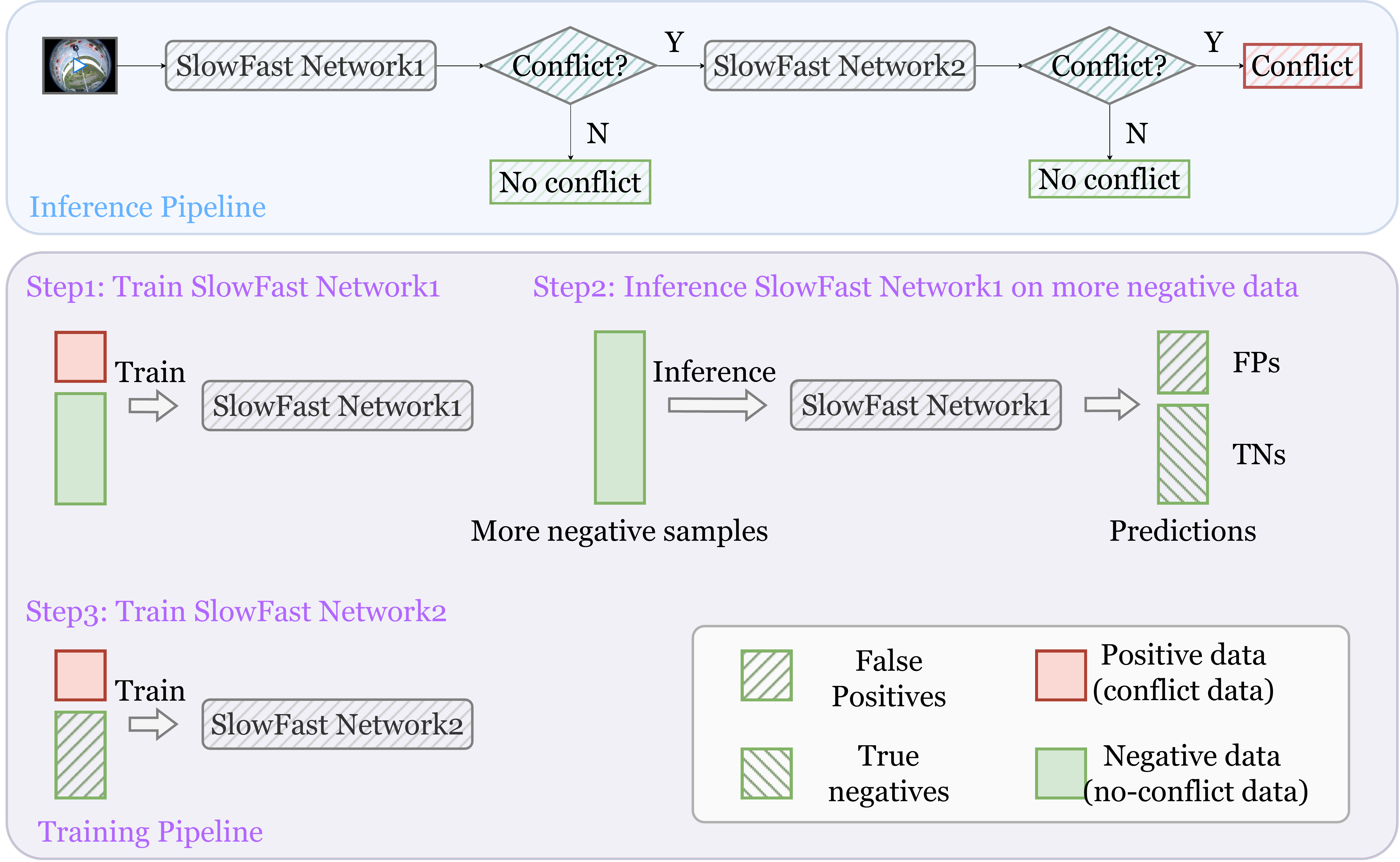}
    \caption{An illustration of the inference
    and training process
    of the cascade conflict identification algorithm.
    The algorithm consists of two SlowFast networks.
    For inference, only when both of two networks predict
    that there is a conflict, the final prediction
    is positive. 
    For training, there are three steps:
    we first train the SlowFast Network1.
    Then we infer the network on more negative samples
    to get the False Positive predictions of SlowFast
    Network1.
    Finally, we use the False Positives as negative
    samples to train SlowFast Network2.
    In this way, the SlowFast Network2 is trained
    to distinguish between real conflicts
    and False Positives of SlowFast Network1.
    }
    \label{fig:cascade-conflict-identification}
\end{figure}

\subsection{Conflict identification algorithm}

At first, we developed a rule-based algorithm to identify
potential conflicts and manually check and annotate them, giving us $232$ initial traffic conflict samples.
Then we use the $232$ traffic conflict samples
to train a learning-based cascade framework for conflict
identification.
As shown in Figure~\ref{fig:cascade-conflict-identification},
there are two
stages in the cascade framework.
The model architecture of the two stages is exactly
the same, both are SlowFast~\cite{feichtenhofer2019slowfast} networks
with binary classification (conflict or not) output.
The SlowFast Network is a commonly-used video processing network.
It takes raw videos as inputs.
There are two pathways in the SlowFast Network:
one slow pathway, operating at low frame rate to
learn spatial features, and one fast pathway,
operating at high frame rate to learn useful
temporal information.
The two networks have the same architecture,
yet different parameters due to different training.
During inference, only when both SlowFast Network1
and SlowFast Network2 predict the sample as a conflict,
we take the final prediction as a conflict.

For training, as shown in Figure~\ref{fig:cascade-conflict-identification},
we first construct a training set
of SlowFast Network1 by assigning conflict videos as
positive data, and randomly sampling negative data.
Then we use this training set to train SlowFast Network1.
Since traffic conflicts are very rare in real-world, 
the precision of the SlowFast Network1 is pretty low ($7\%$).
Most of the predictions of SlowFast Network1 are False Positives,
which means most of the predicted conflict videos actually
have no conflicts.
To reduce the number of False Positives,
we implement another network, SlowFast Network2,
specifically for filtering the False Positives
of SlowFast Network1.
We infer SlowFast Network1 on more negative samples,
and collect the False Positive samples.
Then we construct a training set for SlowFast Network2:
assigning conflict videos as positive data,
and SlowFast Network1's False Positives as negative data.
Thus, the SlowFast Network2 is trained
to tell the difference between real conflicts
and SlowFast Network1's False Positives.

We estimate the precision for both the first-stage
model and the two-stage model. 
The precision is defined as

\begin{equation}
    \quad\quad\quad
    \quad\quad\quad
    \quad\quad\quad
    \quad\quad\quad
    \operatorname{Precision} = 
    \frac{\operatorname{TP}}{\operatorname{TP} + \operatorname{FP}}
\end{equation}

\vspace{4mm}

Here $\operatorname{TP}$ means number of True Positives,
$\operatorname{FP}$ means number of
False Positives, $\operatorname{FN}$ means False Negatives.
Precision is the fraction of real conflicts among all the
detected conflicts.
For the first-stage model,
the precision rate of the output videos is
about $7\%$.
When we feed the output videos of the first-stage
model into the second-stage model for
re-identification,
then the precision rate is improved to $23\%$.
With the help of the traffic conflict
identification model,
the annotation cost is reduced dramatically.

\section{Data Annotation}
\subsection{Traffic conflict annotation and taxonomy}
With the collected potential conflict samples,
we design an annotation instruction for ROCO
and work on annotating the potential conflict samples.
We define three elements for a traffic conflict:
the severity of the conflict,
the reason for the conflict,
and the effect of the conflict on the traffic flow.

\vspace{1mm}
\noindent\textit{Severity.}
For severity, we define three levels of
severity in our conflict dataset.
Level $0$ means that
the data is a negative sample, there
is no traffic conflicts in the clip.
Some vehicles may violate a rule or law,
but it does not place another user in jeopardy
of collision. We do not consider this
type of event as a traffic conflict.
Level $1$ means that there is
one or more traffic conflicts in the clip.
Level $2$ means that a collision 
happens in the clip, there is physical contact
between road users.
We do not define further severity levels,
since we do not find a universal
representation for severity across the whole
dataset, so we only apply the simplest metric.
In the future, other safety metrics
can be applied in the dataset.

\begin{figure}[t]
    \centering
    (a)\includegraphics[width=.21\textwidth,trim={3cm 2.5cm 2.5cm 3cm},clip]{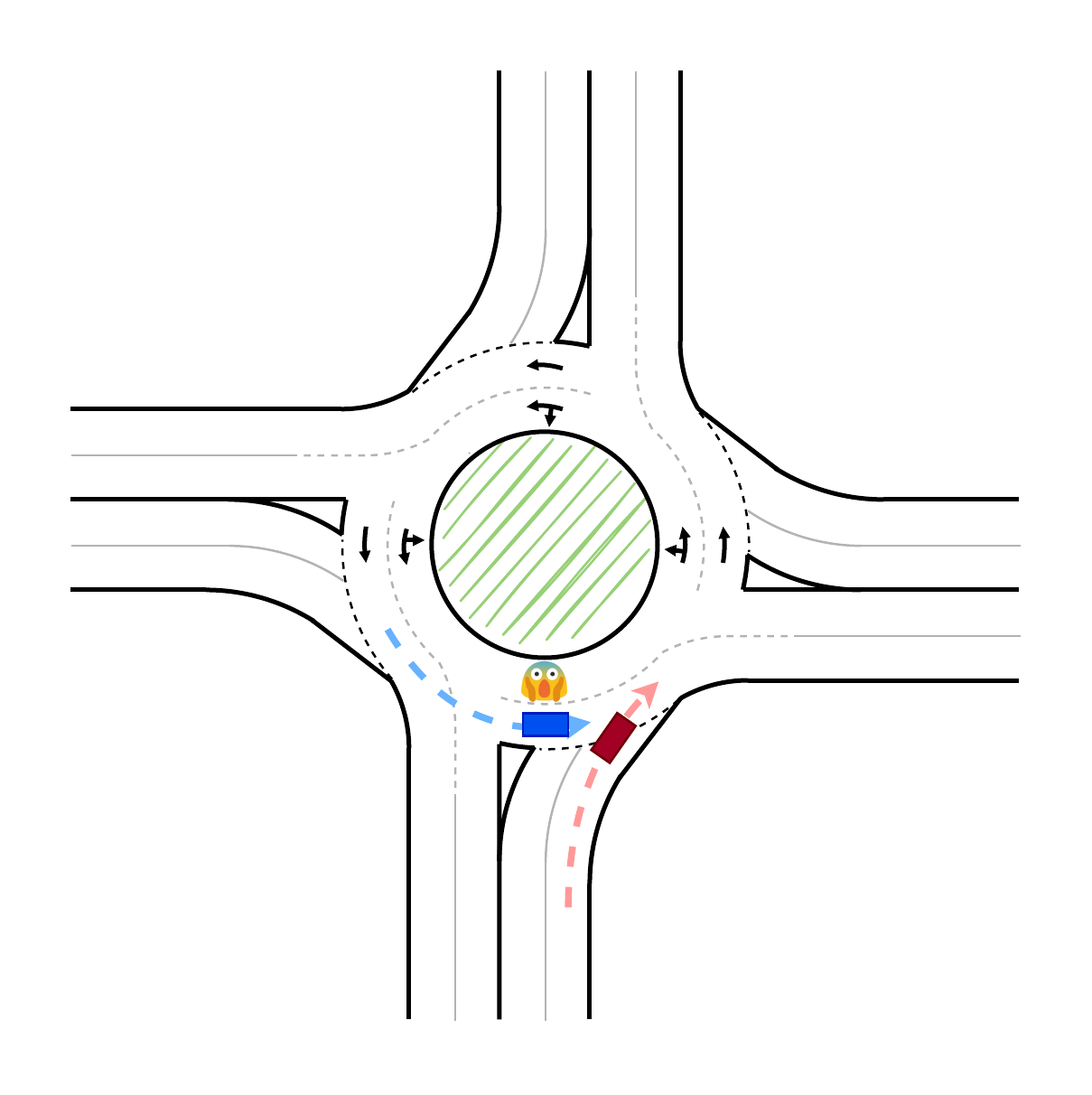}
    (b)\includegraphics[width=.21\textwidth,trim={3cm 2.5cm 2.5cm 3cm},clip]{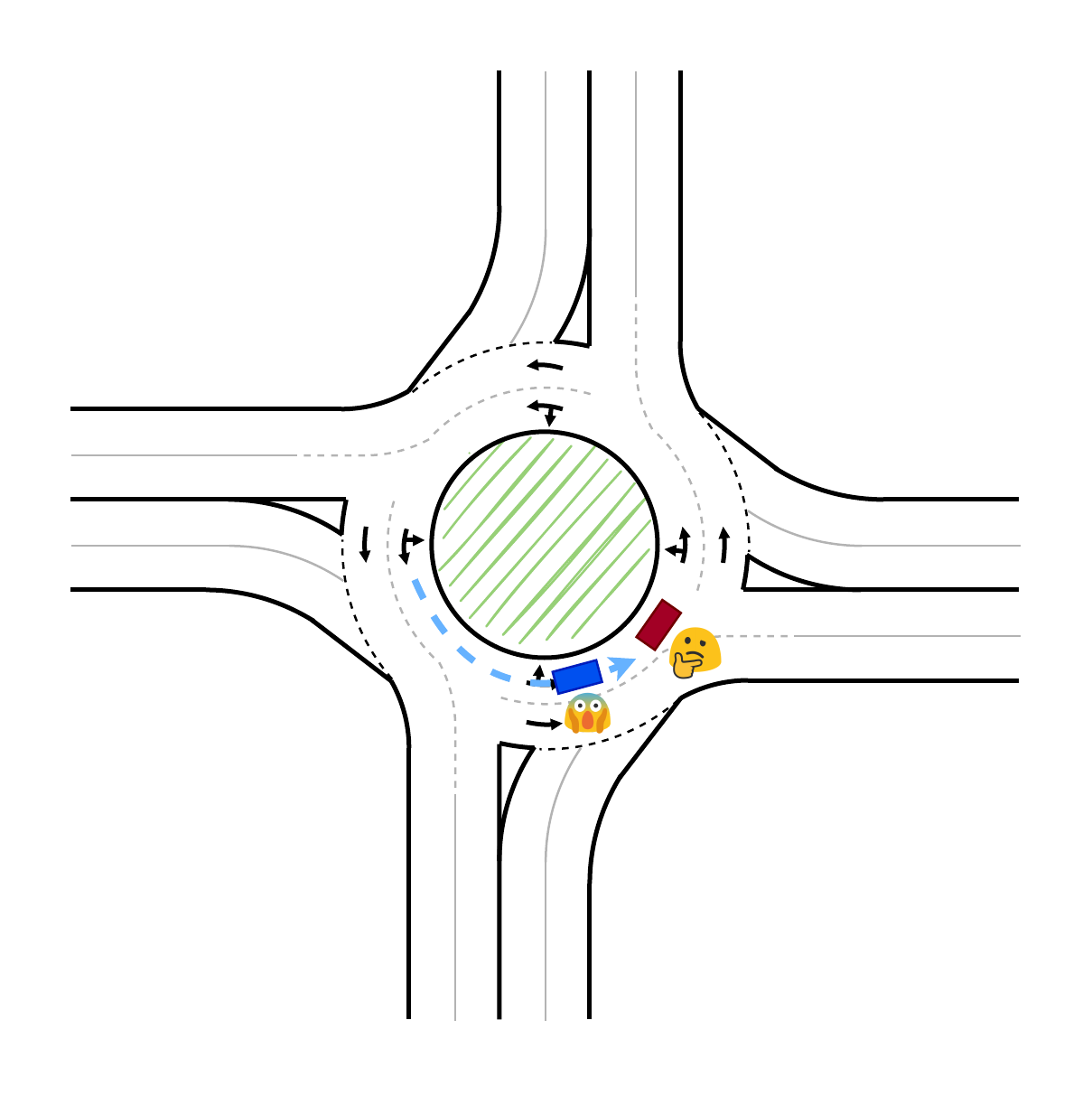}
    (c)\includegraphics[width=.21\textwidth,trim={3cm 2.5cm 2.5cm 3cm},clip]{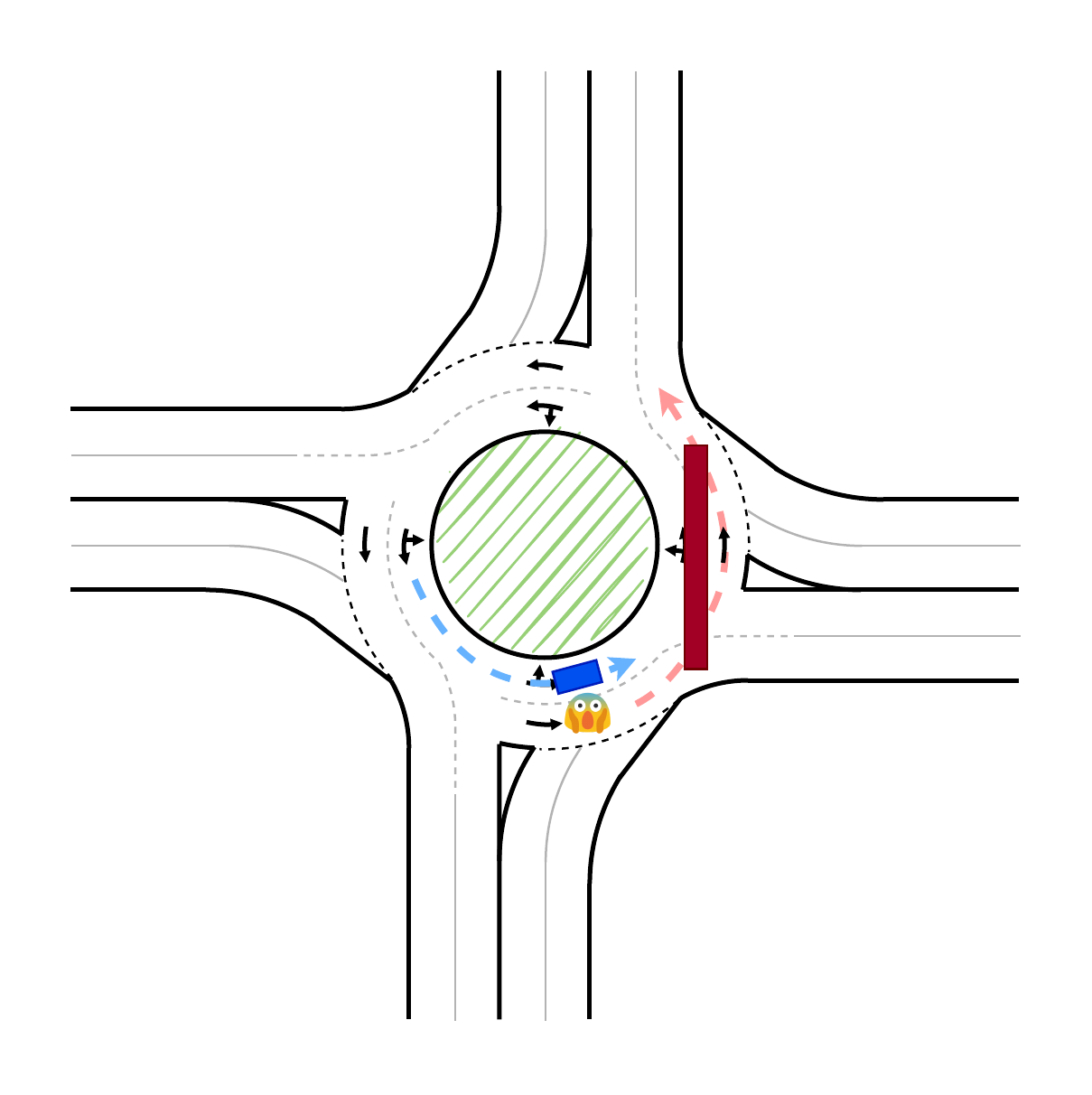}
    (d)\includegraphics[width=.21\textwidth,trim={3cm 2.5cm 2.5cm 3cm},clip]{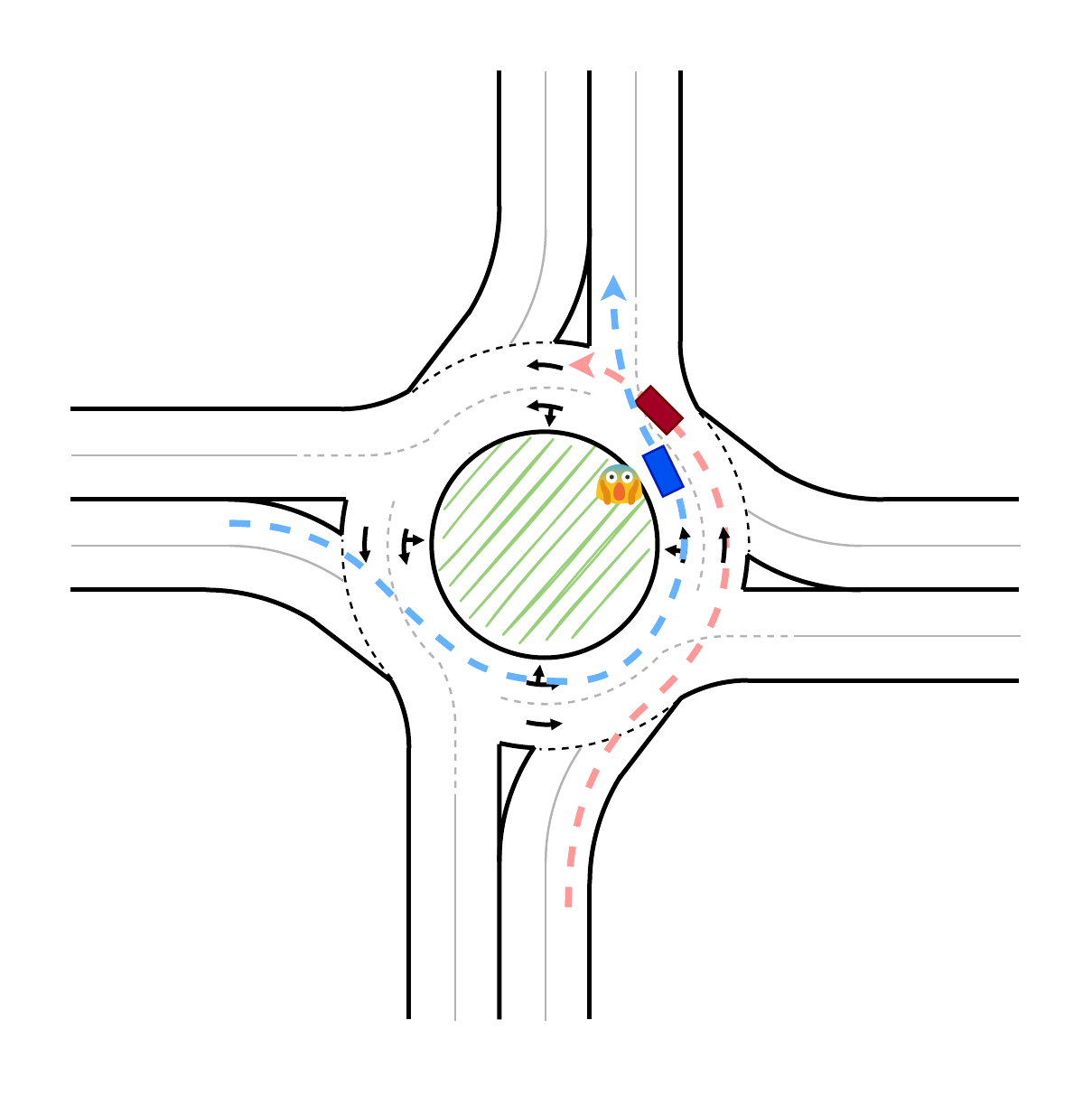}
    \caption{Illustrations of common reasons we summarized
    for traffic conflicts at the roundabout.
    (a) The red vehicle enters the roundabout
    without yielding to the blue vehicle.
    (b) The red vehicle suddenly stops
    in the roundabout and causes a near
    rear-end collision.
    (c) The red truck is too long
    and occupying both lanes of the roundabout
    and force the blue vehicle to slow
    down or stop.
    (d) The red vehicle should exit from the
    top exit but continue to circle in the
    roundabout and causes a near-crash event.}
    \label{fig:traffic-conflict-reasons}
\end{figure}

\vspace{1mm}
\noindent\textit{Reason.}
We divide the reasons for traffic conflicts
in the roundabout to the following common
categories:
\begin{itemize}
\item 
(0) "Entering a roundabout without yielding
to the circulating vehicles".
According to roundabout laws in the state of Michigan, when entering a roundabout, 
the vehicle should yield to traffic circulating
inside the roundabout and wait for a
safe gap to enter.
Failure to do so may cause a safety-critical
event.
An example of failure yield is shown in Figure~\ref{fig:traffic-conflict-reasons}(a).
\item
(0*) "Large vehicles like bus/truck/trailer entering
a roundabout without yielding
to the circulating vehicles". This is a special case of Reason (0),
and in our dataset, it has a higher priority than Reason (0). This means that if a large vehicle
fails to yield and causes a traffic
conflict, then it will be labeled as Reason (0*)
instead of Reason (0).
Since it is harder for large vehicles to enter the
roundabout without interfering with circulating traffic,
we make it a separate reason.
\item
(1) "Unnecessary sudden braking/stop in the circle".
When vehicles are circulating in the circle,
they should not suddenly begin braking or stopping, unless there is an emergency. Sudden braking/stopping
can result in rear-end crashes or near-crashes, as shown
in Figure~\ref{fig:traffic-conflict-reasons}(b).
\item
(2) "Truck/bus/trailer is too large and it interferes with
other road users". Since some vehicles are
too long or wide to comfortably fit in the roundabout, they may take up portions of both lanes.
This can interfere with the other vehicles in the roundabout. Safety critical
events between the large vehicle and surrounding vehicles can form,
as shown in Figure~\ref{fig:traffic-conflict-reasons}(c).
\item
(3) "Improper lane use". Leaving the roundabout
at the wrong exit for a given lane, or crossing
a lane in the roundabout disrupts the
traffic flow, possibly causing a side-swipe collision.
Figure~\ref{fig:traffic-conflict-reasons}(d)
shows an example.
\item
(4) "Other reasons". Many other reasons may
cause a traffic conflict. These include
traffic congestion and police vehicles/ambulances/firetrucks or other emergency vehicles.
We put all these into the other reasons
category.
\end{itemize}

From our observations, the reasons above
cover most of the causes of traffic conflicts
at this roundabout.
For other roundabouts,
the set of common reasons may need to
be extended accordingly in the future.
With these reasons for traffic conflicts,
transportation researchers and government
agencies might be able to better improve the
driving education, traffic signs at roundabouts, etc
to improve traffic safety.

\begin{figure}[t]
    \centering
    \includegraphics[width=.19\textwidth, height=.19\textwidth, trim={13cm 10cm 4cm 7cm}, clip]{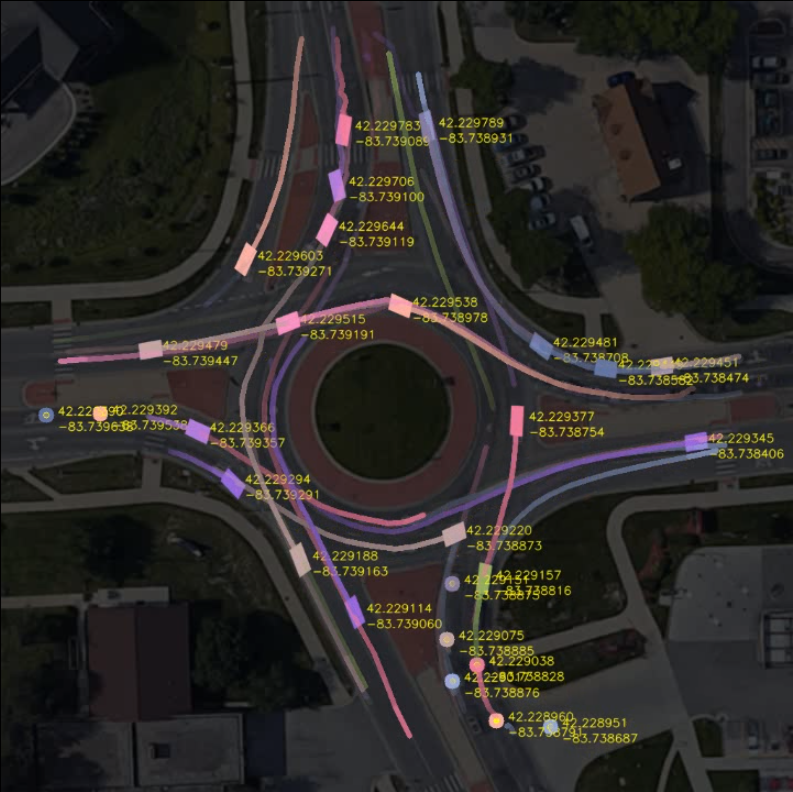}~
    \includegraphics[width=.19\textwidth, height=.19\textwidth, trim={13cm 10cm 4cm 7cm}, clip]{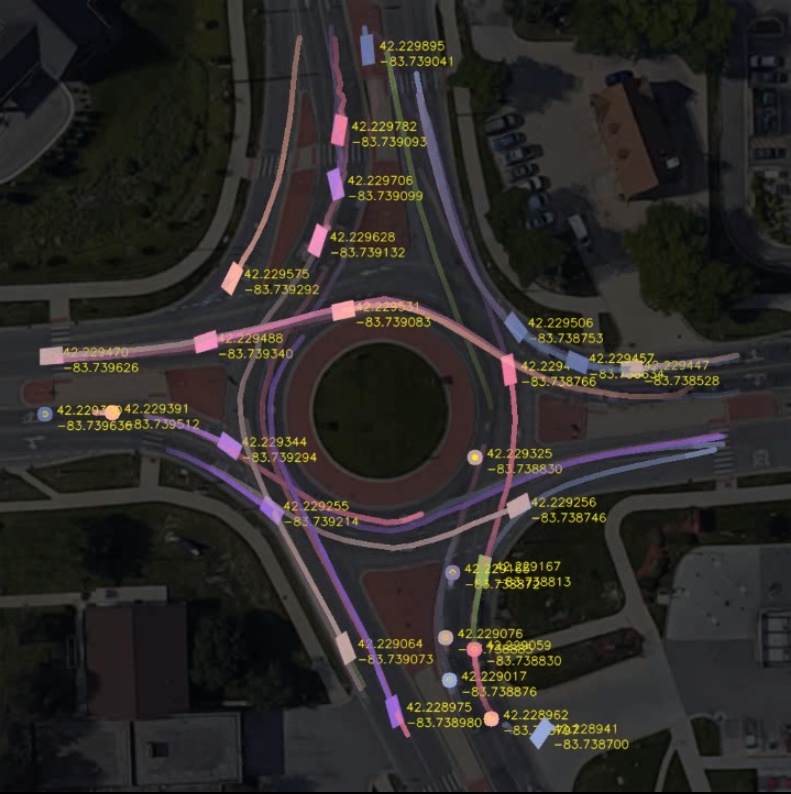}~
    \includegraphics[width=.19\textwidth, height=.19\textwidth, trim={13cm 10cm 4cm 7cm}, clip]{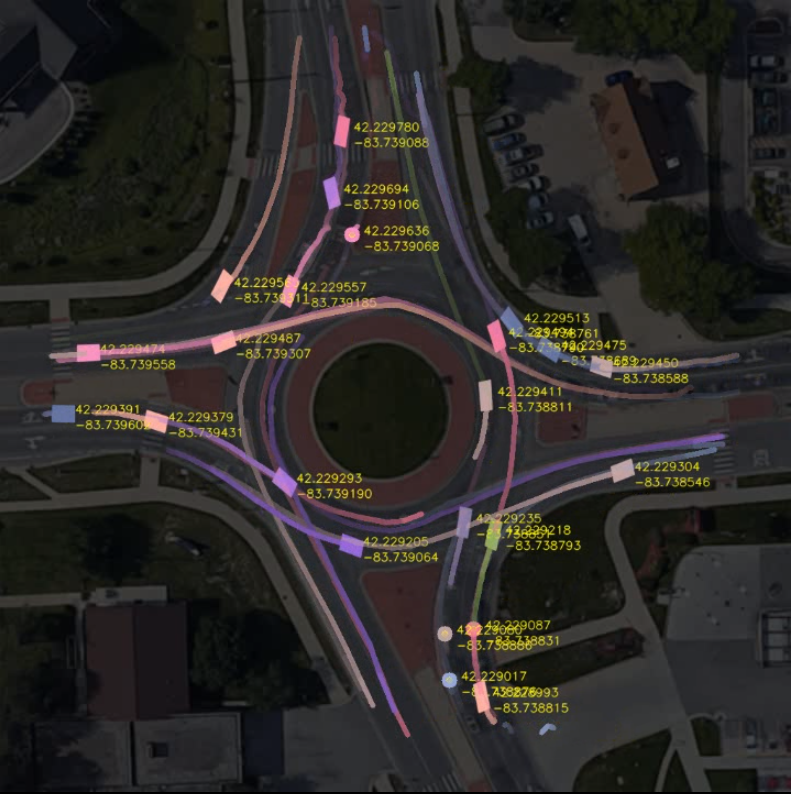}~
    \includegraphics[width=.19\textwidth, height=.19\textwidth, trim={13cm 10cm 4cm 7cm}, clip]{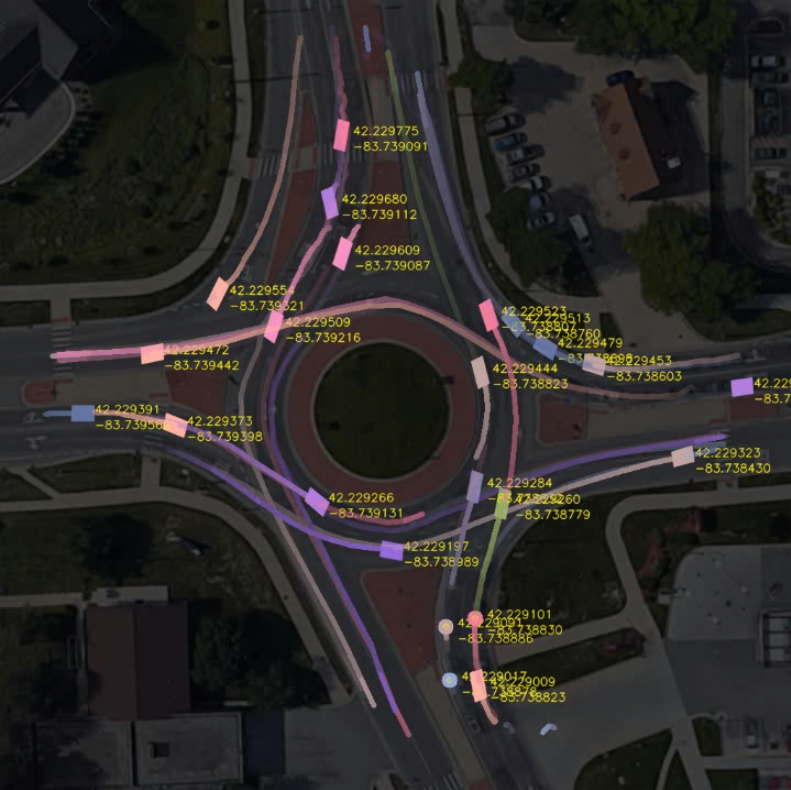}~
    \includegraphics[width=.19\textwidth, height=.19\textwidth, trim={13cm 10cm 4cm 7cm}, clip]{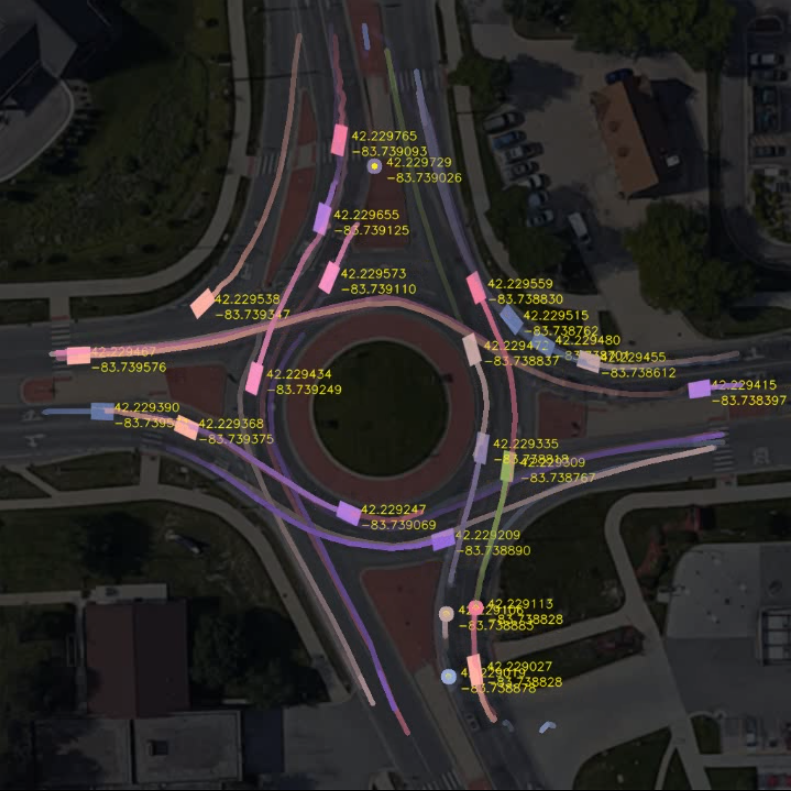}\\
    \caption{Visualizations of extracted trajectories
    of traffic conflicts. The five images are five frames
    during a traffic conflict. The blue vehicle enters
    the roundabout without yielding to the pink
    vehicle causing a traffic conflict.}
    \label{fig:traffic-conflict-trajectory}
\end{figure}

\vspace{1mm}
\noindent\textit{Effect on the traffic flow.}
To better understand how traffic conflicts affect other vehicles,
we annotate the effect of the traffic conflict on traffic flow in the roundabout.
We divide the severity of the interference caused
by the traffic conflict into four categories:
\begin{itemize}
\item 
(0) Traffic flow is not affected. Even the vehicle causing the traffic conflict
does not slow-down or stop. For example, a driver turns the steering wheel sharply
to prevent a crash,
but luckily no vehicle is around so no vehicle
is forced to slow-down.
\item
(1) One or two vehicles are forced to slow down
due to the traffic conflict. This is one common
scenario, since some vehicles slow down to avoid the conflict.
\item
(2) One or two vehicles are forced to stop
in the roundabout due to the traffic conflict.
When the traffic conflict is severe, the driver
might need to brake heavily to avoid the crash.
Vehicles stopping in the roundabout are very dangerous
to other upcoming vehicles.
\item
(3) Three or more vehicles are forced to slow down
or stop in the roundabout.
This means that traffic in the roundabout is
severely affected by the traffic conflict.
\end{itemize}

\subsection{Vehicle detection and tracking system}
Once we identified the video clips
that contain traffic conflicts,
we adopted a vehicle detection and tracking system for trajectory extraction.
We re-use the roadside perception framework
developed by~\cite{Zou2021,Zhang2021},
except we replace the vehicle detector
with a stronger YOLOX~\cite{yolox2021} detector.
With the stronger detector, we are able to
extract more accurate and robust trajectories.
Figure~\ref{fig:traffic-conflict-trajectory} shows an example of the
extracted trajectories for a traffic conflict.

\subsection{Data collection and annotation efficiency}
Currently, with the help of conflict
identification algorithm,
we manage to reduce the data collection and annotation cost
significantly.
The conflict identification algorithm inference speed is
very fast compared to human annotation. With one NVIDIA
RTX $3080$ GPU, it takes around $1$ second inference time
for a $30$ second video clip.
For human annotation, we estimate that for one
video clip, it requires one person $60$ seconds to determine
whether there is a traffic conflict or not (identification),
and an extra $20$ seconds to label the severity,
reason, and effect of the traffic conflict.
Table~\ref{tab:annotation-cost}
illustrates the estimated cost for data collection
and annotation of one traffic conflict data sample
in ROCO.
In ROCO, we estimate that in every $1,200$ video clips,
there will be one video clip containing a traffic conflict,
so if we purely rely on human annotation,
it will cost about $20$ hours human labor
to collect and annotate one traffic conflict.
Fortunately, our traffic conflict identification algorithm
is able to filter out most of the no-conflict
video clips. With an identification precision around $23\%$,
in around $4$ video clips filtered from the identification
algorithm, there is one video clip in it containing a traffic
conflict.
In this case, the estimated human labor cost is reduced
to around $280$s per sample.
If we obtain a perfect traffic conflict identification
algorithm, achieving $100\%$ precision,
the human labor cost will be further reduced, but
still, $80$s per sample is required.

\begin{table}[t]
    \centering
    \small
    \setlength{\tabcolsep}{4pt}
    \begin{tabular}{l|c|c|c|c}
        \shline
        Method & GPU time & Human identification time 
        & Human labeling time & Total labor time \\
        \shline
        Pure human annotation & $0$s & $72,000$s & $20$s & $72,020$s \\
        Our approach & $1,200$s & $260$s & $20$s & $280$s \\
        Perfect identification & $1,200$s & $60$s & $20$s & $80$s \\
        \shline
    \end{tabular}
    \caption{Estimated data collection and annotation cost for
    one traffic conflict data sample in ROCO.
    GPU time means the GPU inference time of our
    traffic conflict identification algorithm.
    Human identification time means
    the time a person needs to determine
    whether there is a conflict in the video or not.
    Human labeling time means
    the time a person needs to label the elements of
    the traffic conflict (severity, reason, etc).
    The total labor time stands for the total
    time required for human labor.
    Perfect identification means an ideal case that
    the conflict
    identification algorithm achieves $100\%$
    precision.
    In this case, the human labor for each sample
    is only $80$ seconds.
    Our approach significantly reduces total
    human labor cost compared to pure human annotation.
    }
    \label{tab:annotation-cost}
\end{table}

\section{Dataset Statistics and Analysis}
This dataset contains $24/7$ video taken from the beginning August of $2021$ until the end of October of $2021$. 
Each video that was flagged with a potential conflict was annotated with information such as whether there was a conflict/crash, reason for the conflict/crash, and the effect of the event on traffic flow.

\begin{figure}[h]
    \centering
    \includegraphics[width=.65\textwidth]{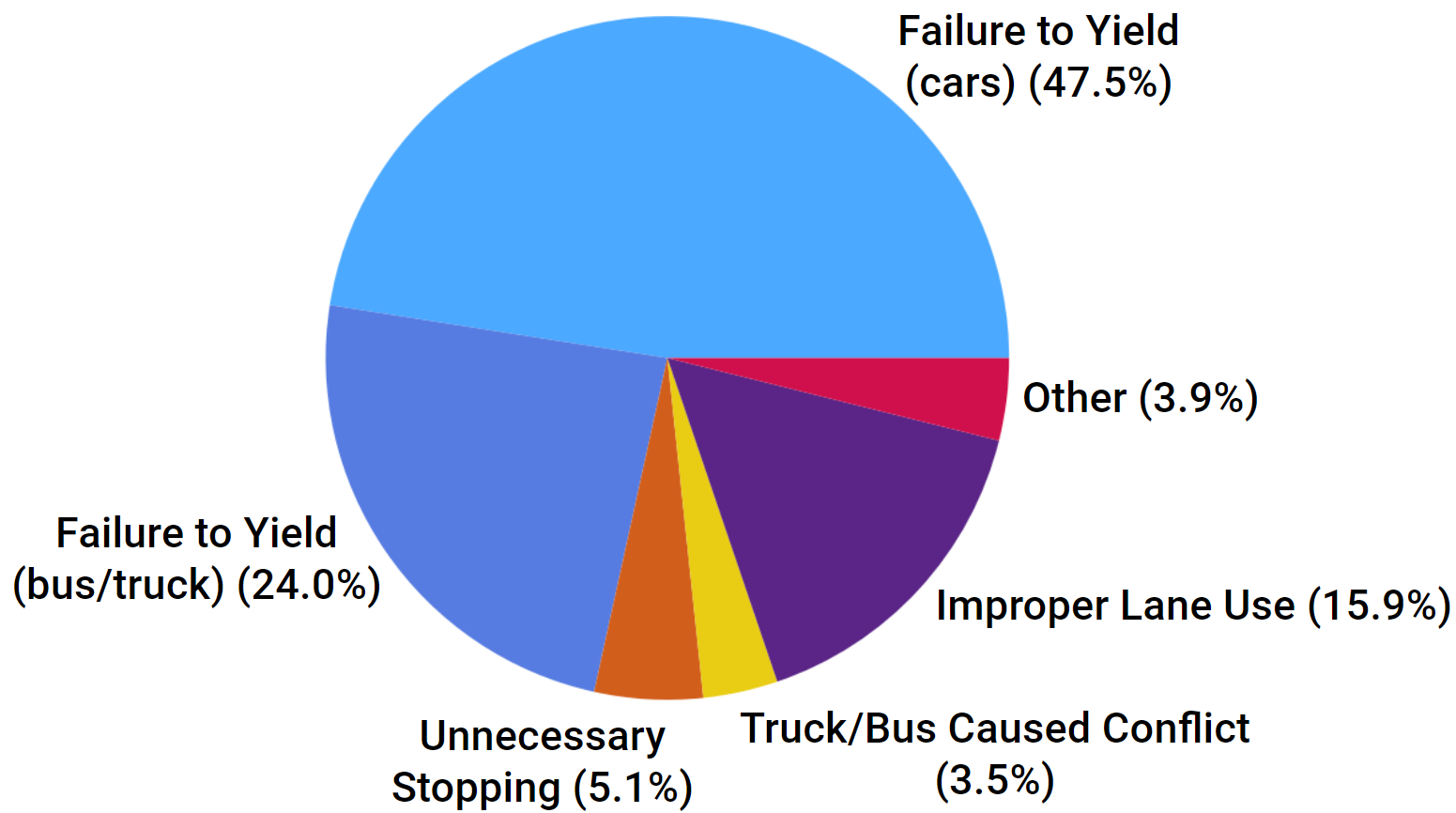}
    \caption{Crashes and Conflicts by Reason. Percentage of conflicts and crashes that were caused by each of the different reasons. Failure to yield is clearly the largest contributor, followed by improper lane use. These show that perhaps drivers are unfamiliar with roundabouts or need more experience with them in a safer or controlled environment.}
    \label{fig:conflict-reasons-pie-chart}
\end{figure}

After having annotated and tagged each potential conflict, some basic statistics were run to analyze the results of the dataset. Overall there were $17$ crashes and $557$ traffic conflicts. Of the conflicts and crashes, the most common reason for the conflict was fail-to-yield to a vehicle already in the circle, followed by failure to yield by a truck, bus, or trailer, and the 3rd most common reason was improper lane use. All of the reasons and what percentage of conflicts and crashes they made up can be seen in Figure~\ref{fig:conflict-reasons-pie-chart}. These are interesting findings because all three reasons are fairly easy to prevent. In the case of failure to yield, perhaps drivers need more experience with roundabouts to learn about and get used to when to enter and when to wait. For improper lane use, drivers may need more education on roundabouts and the lanes, as well as additional or clearer signage at the roundabout.

\begin{figure}[h]
    \centering
    (a)\includegraphics[width=.46\textwidth, height=.25\textwidth, trim={2.5cm 0cm 3cm 4cm},clip]{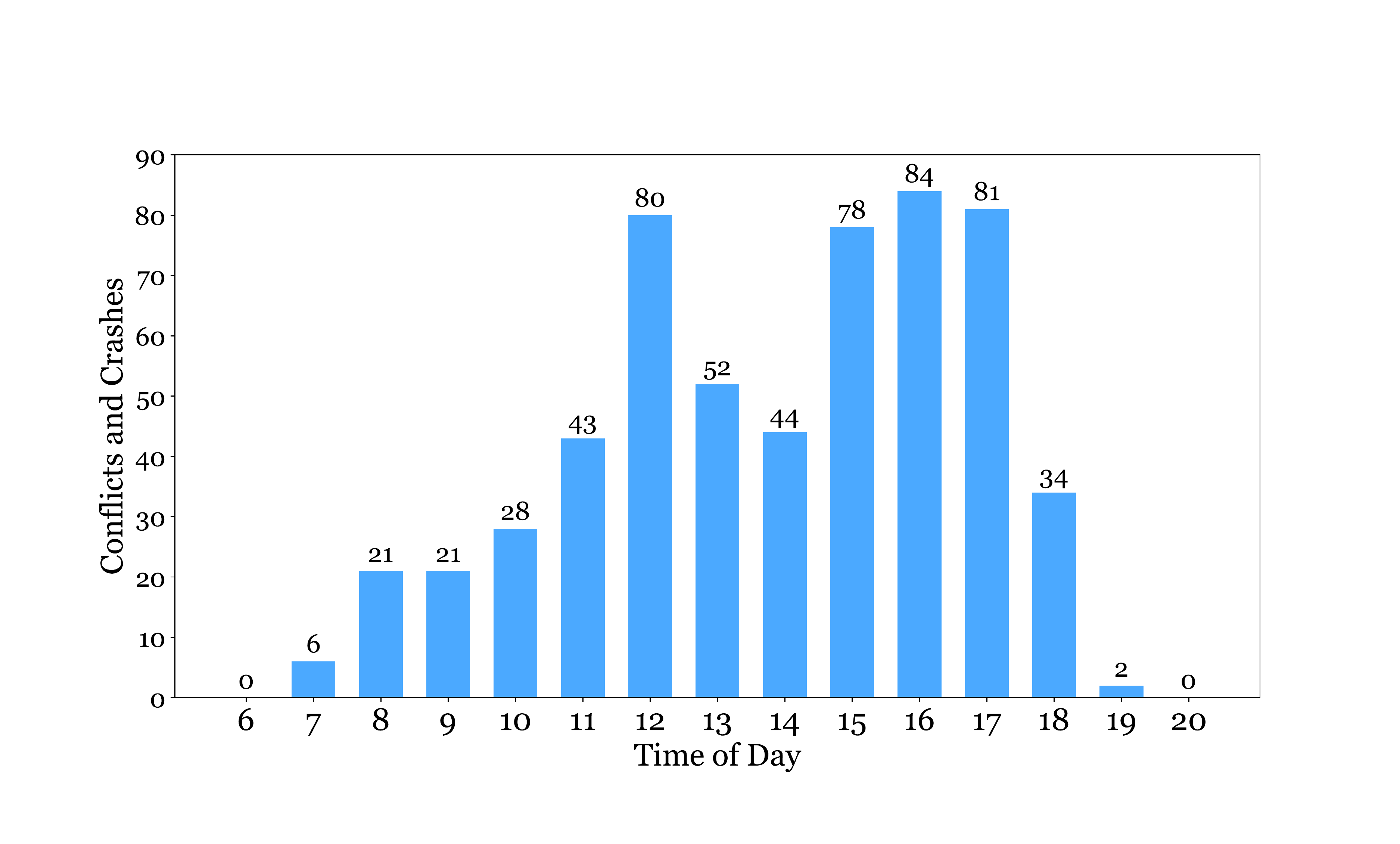}
    (b)\includegraphics[width=0.46\textwidth, height=.25\textwidth,trim={4cm 4.4cm 2.5cm 1.3cm},clip]{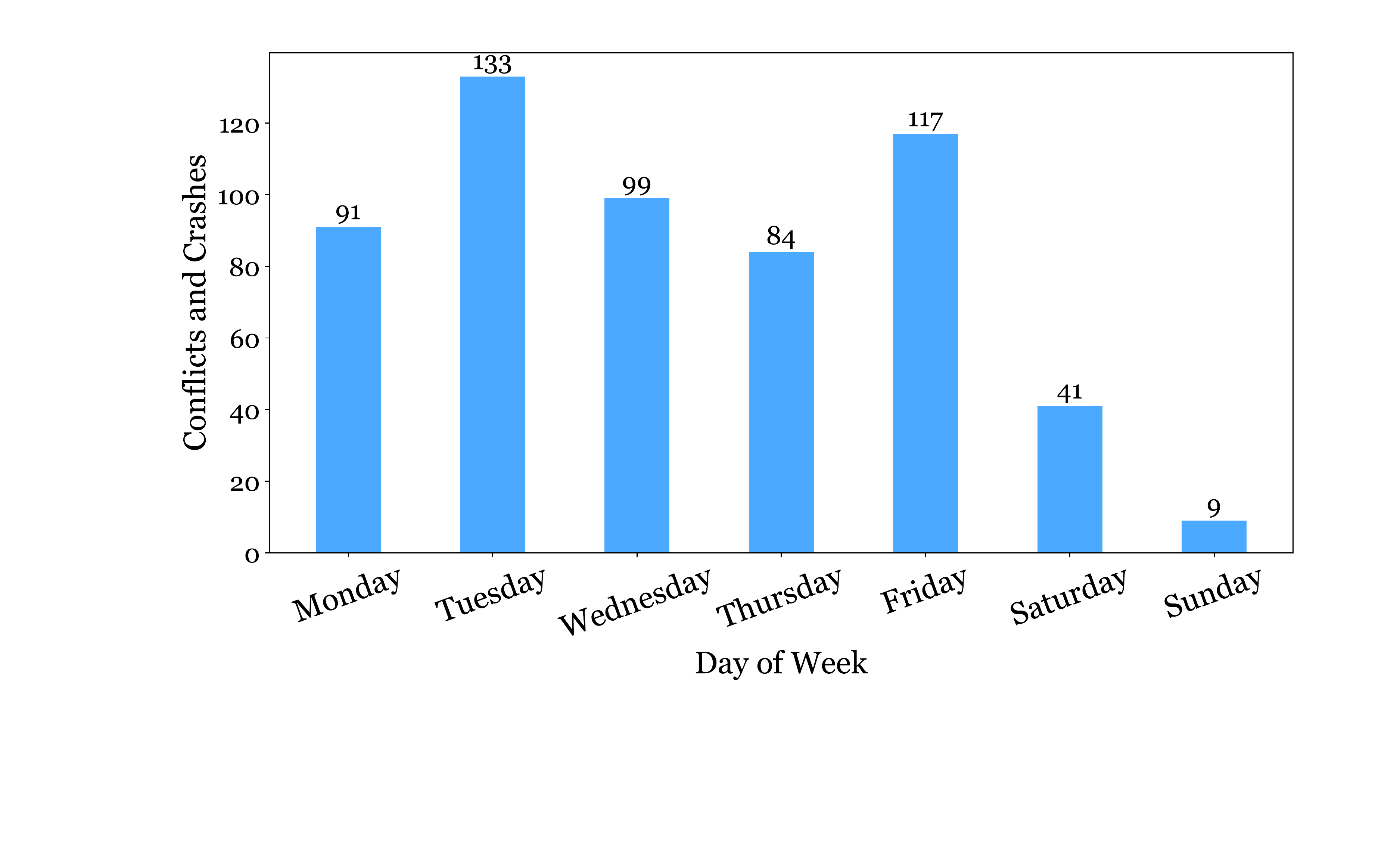}
    \caption{
    (a) Conflicts and crashes grouped by time of day. 
    (b) Conflicts and crashes sorted by day of the week. 
    (a) All crashes and conflicts are shown by hour between 6am and 8pm. The data shows a general upward trend in conflicts as the day goes on, until about 6pm when they start to drop off.
    This can likely be explained, at least in part, by a standard work day which ends around 5pm.
    (b) It's interesting to see that Tuesday has a much larger number of conflicts than do most of the rest of the weekdays do. A fairly consistent number everyday of the work week would seem normal.}
    \label{fig:time-of-day-chart}
\end{figure}

Conflicts and crashes were also separated by hour of the day. This is shown in Figure~\ref{fig:time-of-day-chart}(a). We can see that people tend to crash much more often in the afternoon hours than in the morning. Specifically between 3pm and 6pm; perhaps because of increased traffic in the afternoon. There also seems to be a large number of conflicts around noon. 
Something that is worth noting about this data is that no potential conflicts were flagged between midnight and 6am. This indicates that the camera or program may have trouble detecting vehicles and/or crashes when there is less light, although there are fewer drivers during the night which may partially account for this. Crashes and conflicts were also sorted by day of the week, as displayed in Figure~\ref{fig:time-of-day-chart}(b). One interesting thing about this data is that Tuesday shows a much higher number of conflicts than does most of the rest of the work week. A fairly even number across the days would likely be expected. This may be explained by local events that aren't accounted for.

\begin{figure}[h]
    \centering
    (a)\includegraphics[width=0.45\textwidth]{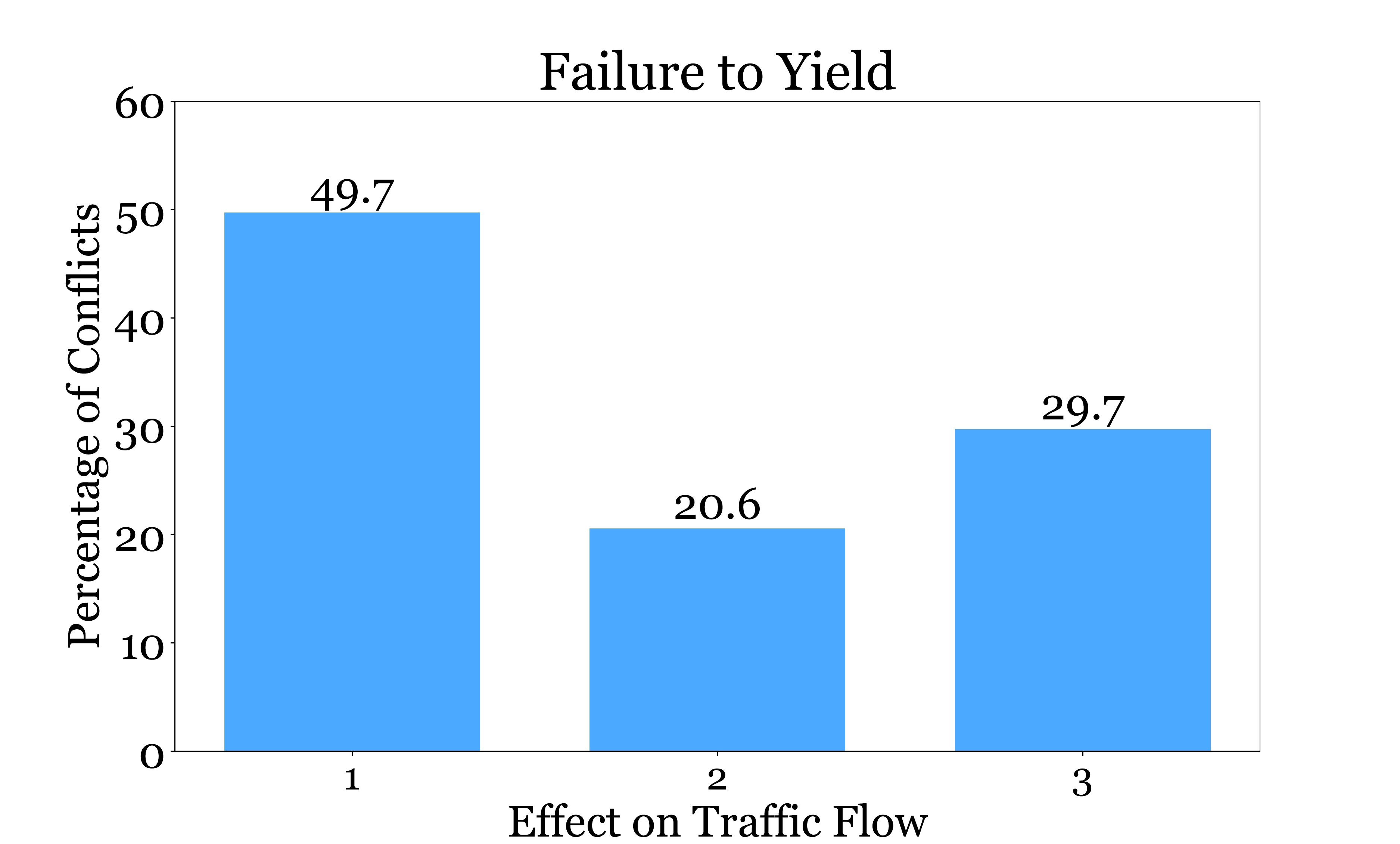}
    (b)\includegraphics[width=0.45\textwidth]{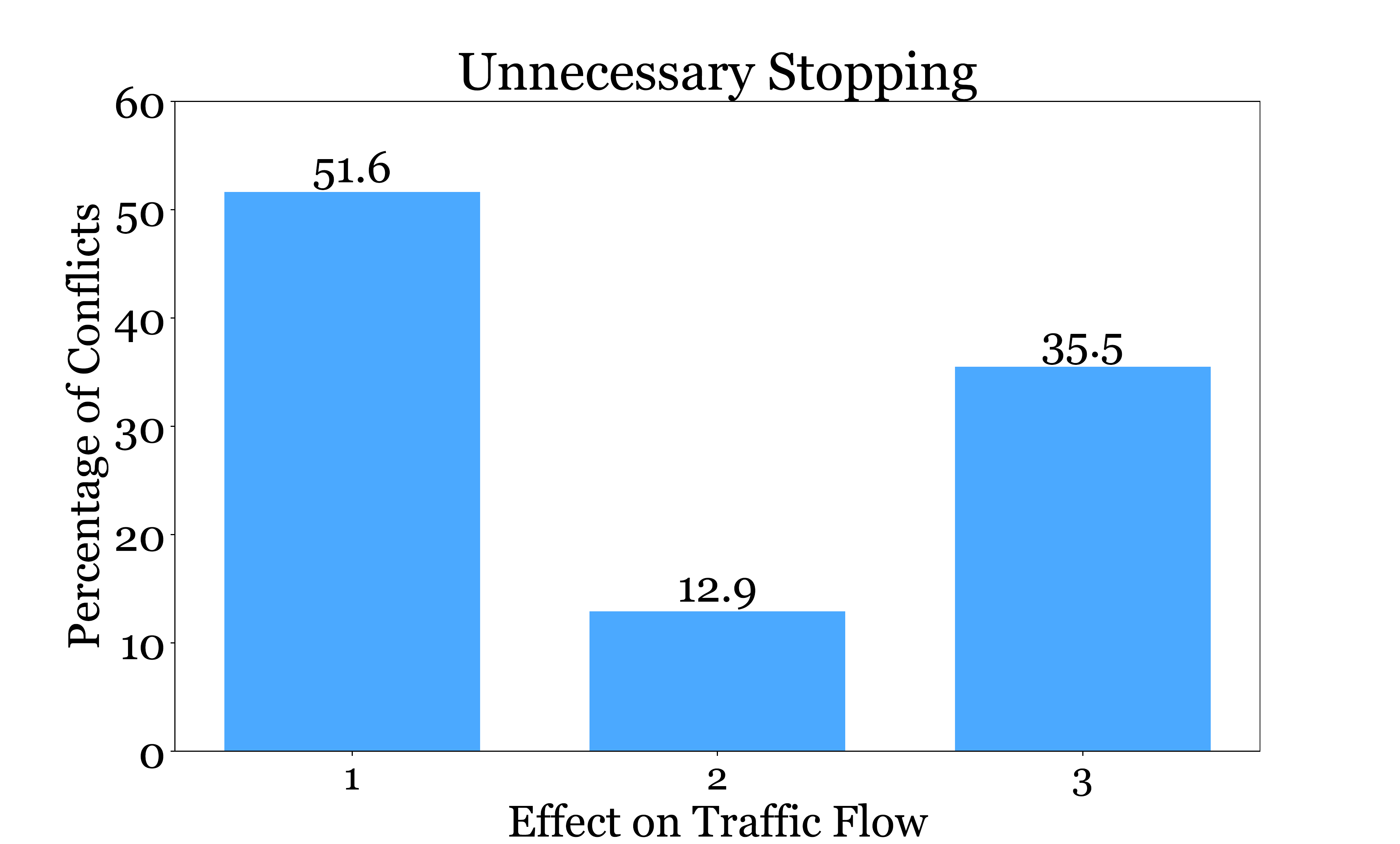} \\
    (c)\includegraphics[width=0.45\textwidth]{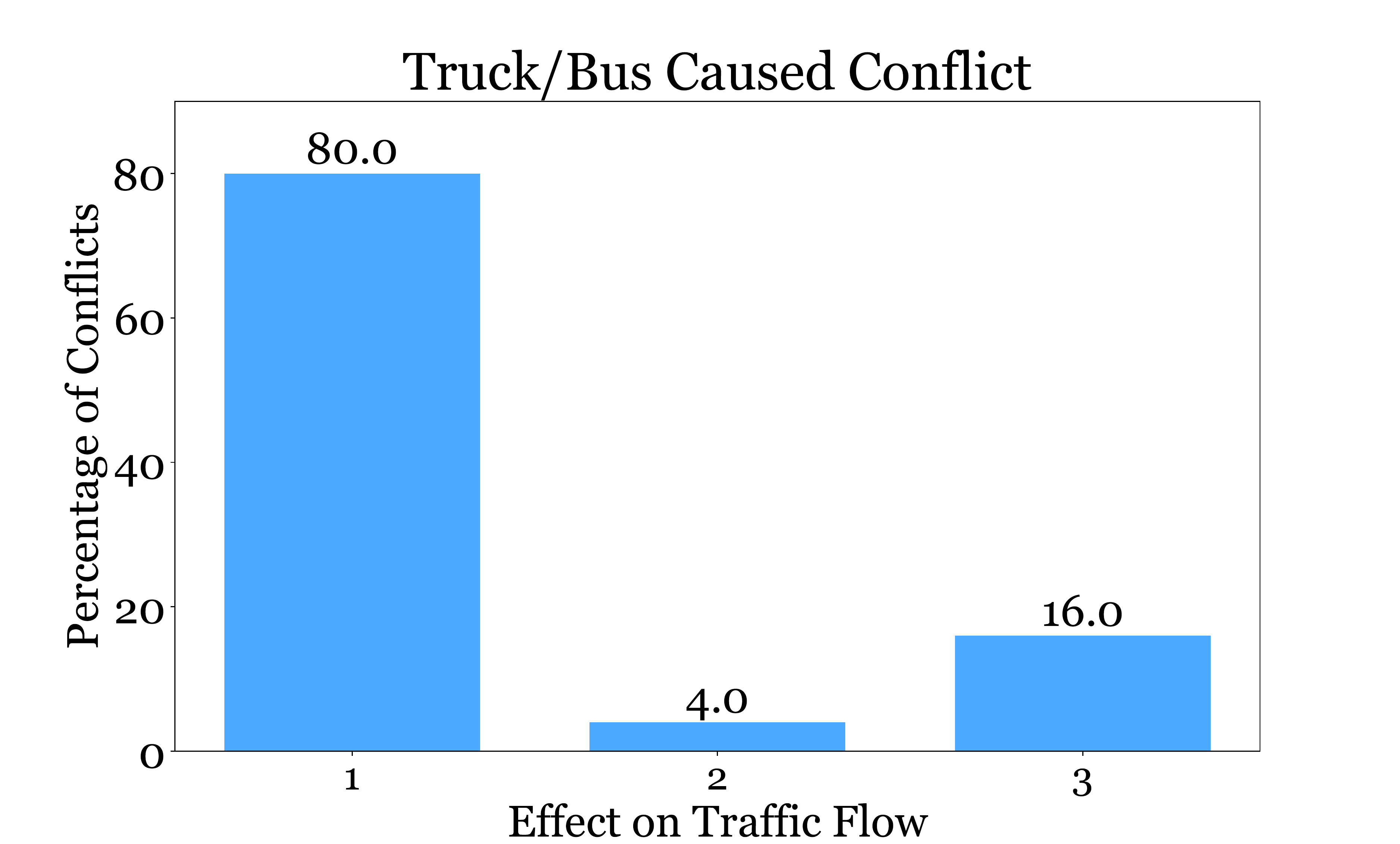}
    (d)\includegraphics[width=0.45\textwidth]{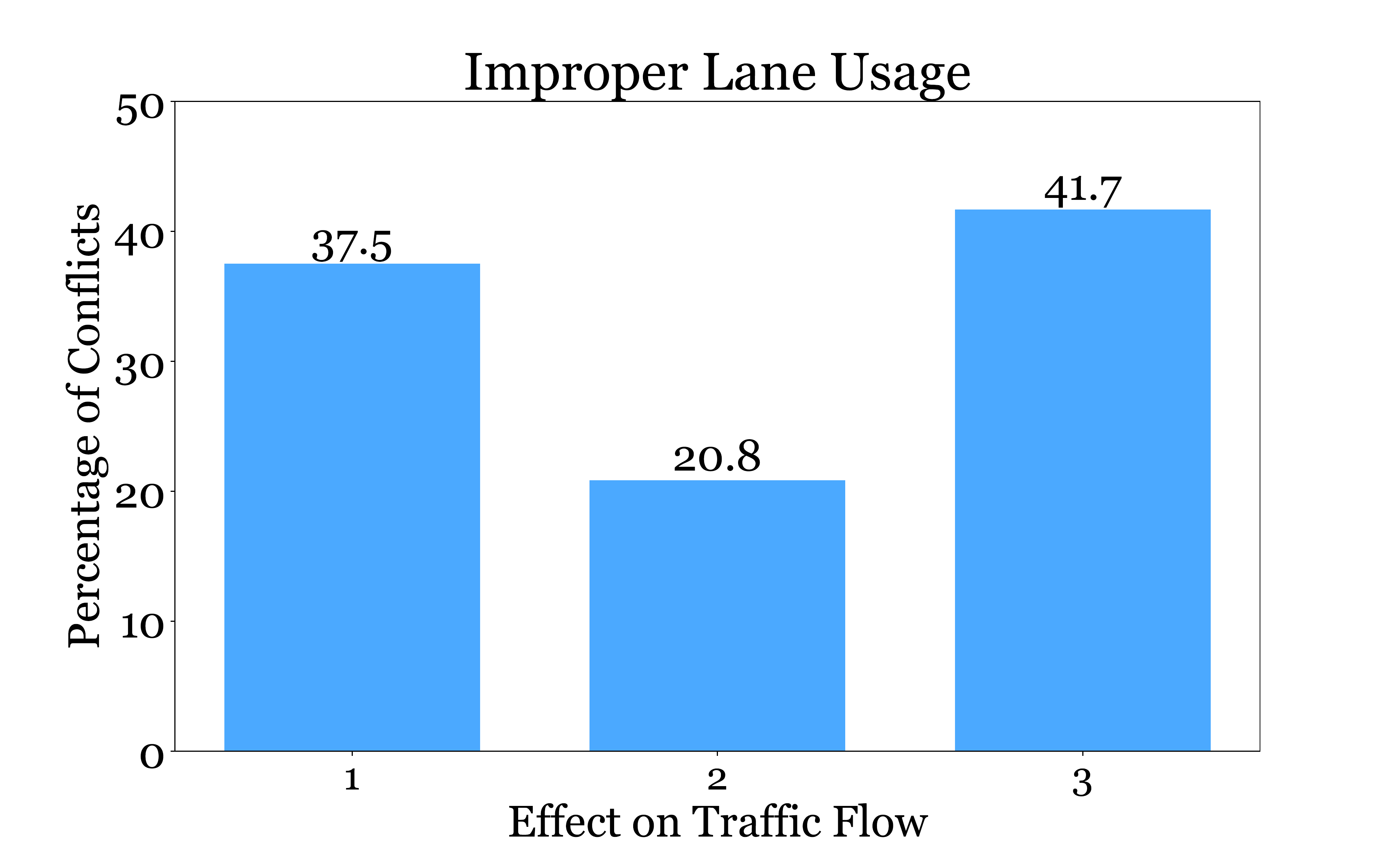}
    \caption{These figures show the percentage of conflicts that a given effect on traffic flow represents for 4 different reasons of conflict: (a) Failure to Yield, (b) Unnecessary Stopping, (c) Truck/Bus Caused Conflict, (d) Improper Lane Usage. 
    The rankings 0-3 represent how severe the effect on traffic flow. So 0 is none, 1 is 1-2 cars slow down, 2 is 1-2 cars stop, 3 is 3 or more cars slow down or stop.
    For example, about 49.7 percent of all failure-to-yield conflicts had a severity of 1, so they resulted in 1-2 cars slowing down.}
    \label{fig:percentage-bar-charts-by-reason}
\end{figure}

In order to better understand the effect on traffic flow of the different reasons for conflicts, several charts were created. 
These are the charts shown in Figure~\ref{fig:percentage-bar-charts-by-reason}. They show the percentage of the conflicts with a given reason that had the given effect on traffic flow severity level.
One interesting takeaway from this figure is that improper lane usage was most likely to result in an effect on traffic flow of severity 3, which means 3 or more cars slowed down or stopped. While all of the other reasons clearly show that the most likely effect on traffic flow was a severity of 1, only one or two cars slowed down. 

Overall the ROCO dataset contains $17$ crashes and $557$ traffic conflicts, the vast majority of which are caused by failure to yield to oncoming vehicles. These conflicts and crashes happen mostly during the afternoon around the 3pm to 6pm period; they also mostly occur on weekdays, with a peak on Tuesday. Most conflicts and crashes result in only a minor effect on traffic flow, while improper lane usage tends to have a more significant effect.

\section{Limitation and Future Work}
There are a few limitations of the ROCO dataset.
The first one is the currently relatively small data size due to the human labor cost; 
The second one is the potential data bias problem
brought by the traffic conflict identification
algorithm we applied. We also discuss the data generalization to more traffic scenarios in this section.

\subsection{Dataset size}
Our current dataset only contains $557$ traffic
conflicts and $17$ crashes.
As discussed in the \emph{Data collection}
section, the human labor cost
of identifying one traffic conflict in ROCO
dataset is about $280$ seconds.
Since traffic conflict is
a very specialized type of event,
we can only trust people with
transportation knowledge for data annotation.
We plan to further lower the human labor cost
of data annotation so that large-scale
traffic conflict data can be collected.

\subsection{Potential data
bias}
Since we use a traffic conflict identification
algorithm to identify potential traffic
conflicts,
it is possible that some data bias
is introduced by the identification algorithm.
For example, the algorithm
might be more likely to identify
traffic conflicts caused by 'fail to yield'
than other traffic conflict reasons.
Since we do not have all the
traffic conflicts at a certain
time period,
we are not able to estimate the data
bias problem now.
We will try to analyze the potential
data bias issue in the future.

\subsection{Generalization to other traffic scenes}
Currently, we only collect traffic
conflict data from one roundabout.
For different roundabouts, or different
intersections, the patterns
and statistics on the traffic conflicts
might be different.
We plan to collect more traffic conflict
data at different intersections and roundabouts
to provide more diversity and coverage to the dataset.

\section{Conclusion and Broader Impact}
In this paper,
we present ROCO, a roundabout traffic
conflict dataset.
Through adopting a traffic conflict
identification algorithm,
we manage to reduce the annotation
cost to an affordable range.
We provide a variety of labels for traffic
conflicts, including severity,
reason, and effect on traffic flow.
Basic statistical analysis is performed
with the data.
We discover that failure to yield when
entering the roundabout is the No. $1$
reason for traffic conflict
in our dataset.

Unlike crash data, where
one crash has to occur in order
to obtain one crash data,
traffic conflict data collection
can be done without actual crashes happening.
With the data collection and annotation
pipeline, we believe that we can
continue to provide more and more real
traffic conflict data to the research
community and government agencies.
More analysis can be done
by transportation researchers
around the world when the ROCO dataset
made public.
Measures can be taken
(e.g., make the traffic signs more clear)
accordingly with the statistical analysis
to prevent traffic conflicts,
as well as potential
crashes in the roundabout.
Further, in autonomous vehicle
development, testing the
autonomous vehicles in the
safety-critical scenarios
is essential for reliability
of autonomous vehicles.
Our ROCO dataset might be able
to serve as a safety-critical
autonomous vehicles testing dataset.

\section{Acknowledgements}
The authors would like to thank Mcity at the
University of Michigan, Ann Arbor
for financial support.
The authors also would like to thank
all the members of Michigan
Traffic Lab at University of Michigan,
Ann Arbor.
Their help in annotating
the traffic conflict data
is indispensable to
the ROCO dataset and this paper.

\section{Author contributions}
The authors confirm their contribution to the
paper as follows:
study conception and design:
Depu Meng, Shengyin Shen, Henry X. Liu;
Data collection:
Depu Meng, Owen Sayer, Rusheng Zhang, Shengyin Shen;
Analysis and interpretation of results:
Depu Meng, Owen Sayer;
draft manuscript preparation:
Depu Meng, Owen Sayer, Rusheng Zhang, Shengyin Shen,
Houqiang Li, Henry X. Liu.
All authors reviewed the results and approved the final
version of the manuscript.

\newpage

\bibliographystyle{trb}
\bibliography{main}
\end{document}